\documentclass[a4paper]{jpconf}
\usepackage{amsmath, amsthm, amssymb, bm}
\usepackage{amsfonts}
\usepackage{amscd}
\usepackage{delarray}
\usepackage{graphicx}
\newcommand{\bra}[1]{\langle #1 \! \mid}
\newcommand{\ket}[1]{\mid \!\! #1\rangle}
\newcommand{\braket}[2]{\langle #1 \!\mid\!\!  #2 \rangle}
\newcommand{\ketbra}[2]{\mid\!\! #1 \rangle \langle #2 \!\mid}
\newcommand{\vel}[1]{\mathbf{\hat{#1}}}

\newcommand{\abs}[1]{\mid \!\! #1 \!\! \mid}
\newcommand{\s}{\mathsf{s}}

\begin{document}

\title{Quantal time asymmetry:
mathematical foundation and physical interpretation}

\author{A Bohm, P Bryant and Y Sato}

\address{CCQS, Department of Physics, University of Texas at Austin, Austin, TX 78712}

\ead{bohm@physics.utexas.edu, pbryant@physics.utexas.edu, satoyosh@physics.utexas.edu}

\begin{abstract}
For a quantum theory that includes exponentially decaying states and Breit-Wigner
resonances, which are related to each other by the lifetime-width relation
$\tau=\frac{\hbar}{\Gamma}$, where $\tau$ is the lifetime of the decaying state
and $\Gamma$ the width of the resonance, one has to go beyond the Hilbert space
and beyond the Schwartz-Rigged Hilbert Space $\Phi\subset\mathcal{H}\subset\Phi^\times$
of the Dirac formalism.
One has to distinguish between prepared states, using a space $\Phi_-\subset\mathcal{H}$,
and detected observables, using a space $\Phi_+\subset\mathcal{H}$, where $-(+)$ refers
to analyticity of the energy wave function in the lower (upper) complex energy semiplane.

This differentiation is also justified by causality: A state needs to be prepared first,
before an observable can be measured in it.
The axiom that will lead to the lifetime-width relation is that $\Phi_+$ and $\Phi_-$
are Hardy spaces of the upper and lower semiplane, respectively.
Applying this axiom to the relativistic case for the variable
$\s=p_\mu p^\mu$ leads to semigroup transformations into the forward light cone
(Einstein causality) and a precise definition of resonance mass and width.
\end{abstract}

\section{Time Asymmetry}
Time Asymmetry does \textbf{not} mean \textit{time reversal non-invariance}, i.e.
it is not
described by a Hamiltonian $H$ that does not commute with the time reversal $A_T$ or the
$CP$ operator \cite{kleinknecht}:
\begin{equation}
[A_T,H]\ne 0\quad\textrm{or}\quad[C\,P,H]\ne 0.
\end{equation}
It may be intriguing, however, to speculate about a possible connection \cite{bohm_96_97}.

Time Asymmetry does \textbf{not} mean \textit{irreversibility} or
Thermodynamic Arrow of Time.  It is not
entropy increase in an isolated classical system, $\frac{d S}{d t}>0$.
There may, however, be  some
relation to entropy increase.
For instance, Peierls (1979) 
explains entropy increase from initial boundary conditions in
Boltzmann's Stosszahl Ansatz \cite{peierls}.

Time Asymmetry also does \textbf{not} mean the usual quantum mechanical Arrow
of Time for ``open'' quantum systems, brought about by an external reservoir \cite{edavis}
described by the Liouville equation,
\begin{equation}
\frac{d W}{d t}=L\,W(t)\equiv-i[H,W(t)]+\mathcal{I}\,W(t),
\end{equation}
in which a ``superoperator,'' $\mathcal{I}$, describes the external effect
of the reservoir, or of a measurement apparatus, or of another environment
(decoherence).

By Time Asymmetry, we mean \textit{time asymmetric boundary conditions}
for time symmetric dynamical equations.

The best known example is 
the Radiation Arrow of Time \cite{pdavis}.
Maxwell's equations (dynamical differential equations) are symmetric
in time.  A boundary condition excludes the strictly incoming fields
and selects only the retarded fields of the other sources in the region:
\begin{equation}
A^\mu(x)=A^\mu_{ret}(x)+A^\mu_{in}(x)=A^\mu_{ret}(x).
\end{equation}
\begin{equation}
A^\mu_{in}(x)=0\quad\textrm{(Sommerfeld radiation condition)}
\end{equation}
The boundary condition is
an additional ``principle of nature'' that chooses of the 
two solutions of the Maxwell equations,
\begin{equation}
A^\mu_{\mp}(\vec{x},t)=\int\,\delta\Big(t^\prime-(t\mp\frac{\vert\vec{x}-\vec{x^\prime}\vert}{c})\Big)\,
\frac{j^\mu(\vec{x^\prime},t^\prime)}{\vert \vec{x}-\vec{x^\prime} \vert}\, 
d^3x^\prime d t^\prime,
\end{equation}
the retarded solution
\begin{equation}
A^\mu_{ret}(\vec{x},t)\equiv A^\mu_-(\vec{x},t)=\int
\frac{j^\mu (\vec{x^\prime},t-\frac{\vert\vec{x}-\vec{x^\prime}\vert}{c})}
{\vert \vec{x}-\vec{x^\prime} \vert} d^3 x^\prime.
\end{equation}
The ``disturbance'' $A^\mu(x)$ at the position $\vec{x}$ 
at time $t$ is caused by the source
$j^\mu$ at another point $\vec{x^\prime}$, 
at an earlier time 
$t^\prime=t-\frac{\vert \vec{x}-\vec{x^\prime} \vert}{c}\leq t$.
Radiation must be emitted (at $t^\prime$) 
by a source before it can be detected
by a receiver at $t\geq t^\prime$.

\section{Dynamical Equations and Their Boundary Conditions}

Standard quantum mechanics lacks time asymmetric boundary conditions.
The dynamical equations are\\[5pt]
\begin{minipage}{0.45\textwidth}
$\qquad$\textit{In the Heisenberg picture}
\begin{subequations}
\begin{eqnarray}
\label{heis_eqn}
i\hbar\frac{\partial \Lambda(t)}{\partial t}=-[H,\Lambda(t)]\\
i\hbar\frac{\partial}{\partial t}\psi(t)=-H\psi(t) 
\label{eq:heis_eq2}
\end{eqnarray}
\label{eq:heis_eom}
\end{subequations}
for the observable $\Lambda(t)=\ketbra{\psi(t)}{\psi(t)}$, \\
with state $W=\ketbra{\phi}{\phi}$ kept fixed.
\end{minipage}\hspace*{22pt} or\hspace*{14pt}
\begin{minipage}{0.45\textwidth}
$\qquad$
\textit{In the Schr\"odinger picture}
\begin{subequations}
\begin{eqnarray}
i\hbar\frac{\partial W(t)}{\partial t} = [H, W(t)]
\label{eq:density_eom}
\\
i\hbar\frac{\partial}{\partial t}\phi(t)=H\phi(t)
\label{schrod_eqn}
\end{eqnarray}
\label{eq:schrod_eom}
\end{subequations}
for the state $W(t)$ or $\phi(t)$
with observable $\Lambda=\ketbra{\psi}{\psi}$ kept fixed
\end{minipage}
\\
\\
\noindent
In standard quantum mechanics \cite{von_neumann}, one solves these equations
under the \textit{boundary conditions}
\begin{equation}
\textrm{set of states }\{\phi\}=\mathcal{H}=\textrm{Hilbert space}=
\textrm{set of observables }\{\psi\}.
\label{eq:hilbert_bc}
\end{equation}

\noindent
As a consequence of the Hilbert space boundary condition,
one obtains from the
Stone-von Neumann Theorem \cite{stone_von_neumann}\\
\begin{center}
\begin{minipage}{0.45\textwidth}
\begin{center}
\begin{equation}
\psi(t)=\text{e}^{iHt}\psi,\quad-\infty<t<+\infty
\end{equation}
\emph{in the Heisenberg picture}
\end{center}
\end{minipage}\hspace*{15pt}
\begin{minipage}{0.45\textwidth}
\begin{center}
\begin{equation}
\phi(t)=\text{e}^{-iHt}\phi,\quad-\infty<t<+\infty
\end{equation}
\emph{in the Schr\"odinger picture}
\end{center}
\end{minipage}
\end{center}$\quad$
\\

The sets of operators
\begin{equation}
\label{set1}
\{U(t)=\text{e}^{iHt}:\quad-\infty<t<+\infty\}
\end{equation}
and
\begin{equation}
\label{set2}
\{U^{\dagger}(t)=\text{e}^{-iH^{\dagger}t}:\quad-\infty<t<+\infty\}
\end{equation}
form a group, because the products
$U(t_{1})U(t_{2})=U(t_{1}+t_{2})$
and the inverses $U^{-1}(t)=U(-t)$ exist.
For every observable $\psi(t)=U(t)\psi$ 
at time $t$
there is also an observable $\psi(-t)= U^{-1}(t)\psi$ 
at time $-t$.
The same applies for the state $\phi$ and the operator $U^\dagger(t)$.

In quantum physics one distinguishes \cite{kraus} between
\emph{states}, which are described by density operators $W$ or by  state
vectors $\phi$, and
\textit{observables}, which are described by operators $A=A^{\dagger}$,
$\Lambda=\Lambda^{2}$, or by observable vectors $\psi$ if
$\Lambda=\vert\psi\rangle\langle\psi\vert$.\\[5pt]
\begin{minipage}{0.35\textwidth}
  State $W$ (\textit{in-}states $\phi^{+}$ of scattering experiment.)
\end{minipage}\hfill
\begin{minipage}{0.32\textwidth}
\begin{center}
\emph{is prepared by}
\end{center}
\end{minipage}\hfill
\begin{minipage}{0.32\textwidth}
a preparation apparatus (e.g. accelerator)
\end{minipage}\\[10pt]
\begin{minipage}{0.35\textwidth}
Observable $A$ (\textit{out-}observables $\psi^{-}$, ``out-state'')
\end{minipage}\hfill
\begin{minipage}{0.32\textwidth}
\begin{center}
\emph{is registered by}
\end{center}
\end{minipage}\hfill
\begin{minipage}{0.32\textwidth}
a registration apparatus (e.g. detector)
\end{minipage}\\[15pt]
Experimental quantities, $\mathcal{P}_W(\Lambda(t))$, are the probabilities to measure the observable
$\Lambda$ in the state $W$.
They are calculated in theory as Born Probabilities.
They are measured as ratios of large numbers of detector counts (``relative frequencies'').
\begin{eqnarray}
\mathcal{P}_W(\Lambda(t))\equiv\Tr(\Lambda(t)\,W_0)&=&\Tr(\Lambda_0\,W(t))\approx N(t)/N 
\label{eq:born_prob1}
\\
|\braket{\psi(t)}{\phi}|^2 &=& |\braket{\psi}{\phi(t)}|^2 \\
\textrm{in the Heisenberg picture} & & \textrm{in the Schr\"odinger picture} \nonumber
\end{eqnarray}\\
The comparison between theory and experiment is given by
\begin{equation}
\mathcal{P}_W(\Lambda(t))\approx\frac{N(t)}{N}.
\end{equation}
The left hand side is the calculated prediction, and the right hand side is the ratio of
detector counts, where $N(t)$ and $N$ are ``large'' integers.
The comparison between theory and experiment is indicated by $\approx$.\\

What is the experimental evidence for this time evolution
\emph{group}?
It is obvious that
a state $\phi$ must be prepared before the observable $\ketbra{\psi(t)}{\psi(t)}$ can be measured in it (causality), e.g.
the detector cannot register the decay products before the decaying state has
been prepared.
This means that we have a \textit{Quantum Mechanical Arrow of Time}:

The Born probability to measure the observable $\ketbra{\psi(t)}{\psi(t)}$
in the state $\phi$,
\begin{equation}
\label{born_arrow}
\frac{N(t)}{N}\approx\mathcal{P}_\phi(\psi(t))=\vert \braket{\psi(t)}{\phi} \vert^2
=\vert \braket{e^{i H t}\psi}{\phi} \vert^2
=\vert \braket{\psi}{e^{-i H t}\phi} \vert^2
=\vert \braket{\psi}{\phi(t)} \vert^2,
\end{equation}
\textit{exists (experimentally) only for} $t\geq t_0(=0)$,\\[5pt]
where $t_0$ is the preparation time 
of the state $\phi$.
In contrast,
the unitary group of the Hilbert space axiom predicts
$\vert \braket{\psi(t)}{\phi} \vert^2$
for all $-\infty < t < +\infty$.

As a consequence of this obvious phenomenological condition (causality), one obtains
the Quantum Mechanical Arrow of Time:\\[5pt]
In the Heisenberg picture the time translated observables
\begin{equation}
\psi(t)=e^{i H t}\psi\quad\textrm{are physically defined only for }t>t_0=0.
\end{equation}
In the Schr\"odinger picture the time evolved states
\begin{equation}
\phi(t)=e^{-i H t}\phi\quad\textrm{are physically defined only for }t>t_0=0.
\end{equation}
The time evolution is asymmetric, $0\leq t<\infty$, and given
by the semigroup
\begin{equation}
\label{state_semigroup}
\mathcal{U}^\times(t)=e^{-i H^\times t}\quad\textrm{with }0\leq t<\infty\textrm{ for the states }\phi\textrm{ or }W
\end{equation}
or by the semigroup
\begin{equation}
\label{obs_semigroup}
\mathcal{U}(t)=e^{i H t}\quad\textrm{with }0\leq t<\infty\textrm{ for the observables }\psi\textrm{ or }\Lambda.
\end{equation}
Therefore we have the task:
Find a theory
(for instance choosing new boundary conditions to replace the Hilbert space axiom)
for which the solutions of the Schr\"odinger equation
are given by the 
\textit{semigroup} $\mathcal{U}^\times(t)$ (\ref{state_semigroup}),
and for which the solutions of the Heisenberg equation
are given by the 
\textit{semigroup} $\mathcal{U}(t)$ (\ref{obs_semigroup}).

\noindent
\textbf{Remark: Semigroup symmetries of space-time}

In standard quantum mechanics, time evolution is a subgroup of the space-time symmetry
transformations, which, according to the Wigner-Bargmann theorem,
are represented by the (projective) unitary representations in $\mathcal{H}$ of the
Galilei group $G$ (for non-relativistic space-time) and the Poincare
group $\mathcal{P}$ (for relativistic space-time)~\cite{wigner_52}.

In the RHS formulation~\cite{roberts_66,bohm_boulder,antoine_69} using the Schwartz space Gelfand triplet,
$\Phi\subset\mathcal{H}\subset{\Phi^\times}$,
with the Dirac basis vector expansion
\begin{equation}
\phi=\int d\lambda \ket{\lambda}\braket{\lambda}{\phi}\textrm{ for }
\phi\in\Phi,\textrm{ with }\ket{\lambda}=\ket{\lambda_1, \lambda_2, \lambda_3\ldots\lambda_n}\in\Phi^\times,
\end{equation}
symmetry transformations $g\in G$ are represented by a triplet of operators~\cite{antoine_69,antoine_bohm,antoine_07}\\
\begin{minipage}{0.30\textwidth}
$\quad$ \\
$\quad$ \\
$\quad$ \\
acting on \\
e.g. for time translations
\end{minipage}
\begin{minipage}{0.60\textwidth}
\begin{eqnarray*}
U_\Phi(g^{-1}) \subset & U(g^{-1})=U^\dagger(g) & \subset U^\times(g) \\
\Phi \subset & \mathcal{H}\quad\mathcal{H} & \subset \Phi^\times \\
e^{-i H t}\vert_\Phi \subset & e^{-i H t} = (e^{i H t})^\dagger & \subset e^{-i H^\times t},\quad -\infty<t<\infty.
\end{eqnarray*}
\end{minipage}\\[5pt]
Here $U(g)$ is a unitary representation of $g$ in $\mathcal{H}$.
$U_\Phi(g^{-1})$ is the restriction of $U(g^{-1})$ to the 
dense subspace $\Phi\subset\mathcal{H}$, which 
is a continuous operator with respect
to the $\Phi$ topology, and $U^\times(g)$ is the conjugate operator of $U_\Phi(g^{-1})$
in $\Phi^\times$, defined by
\begin{equation}
\bra{\phi}U^\times(g)\ket{\lambda}=\braket{U_\Phi(g^{-1})\phi}{\lambda},\quad\textrm{for all } 
g\in G,\phi\in\Phi,\ket{\lambda}\in\Phi^\times.
\end{equation}
The algebra of observables $H$, $J_i$, $P_i$, etc is obtained by deriving $U(g)$ using the limits
with respect to the $\Phi$-topology;
they are continuous operators in $\Phi$. 
Their conjugate operators
$H^\times$, $J_i^\times$, etc are 
continuous in $\Phi^\times$~\cite{antoine_bohm,wick_02}.

This is a beautiful theory~\cite{roberts_66,antoine_69,antoine_bohm,antoine_07}, 
but it assumes that for every transformation $g\in G$ of the observable
relative to the state,
$g:\psi\rightarrow\psi^g$,
there exists an inverse transformation \textit{also of the observable relative to the state}
(not the state relative to the observable),
$g^{-1}:\psi\rightarrow\psi^{g^{-1}}$.
For rotations and boosts this is meaningful, but for time translation,
it would give an answer to the question: what was the probability of an observable
$\psi$ in a state at a time $t_0+t$, with $t<0$, before the state will be
prepared at $t_0$?
This contradicts causality.
Therefore one must find a space $\Phi_+$ for the observables $\{\psi\}$ such that
the Galilei transformations of non-relativistic space-time 
are represented by a semigroup
$U_+(R,t,\textbf{x},\textbf{v}),\quad t\ge 0$.
Similarly, for the relativistic case mentioned in Section \ref{application_section},
one must find a space (which we will also call $\Phi_+$) in which
the transformations of the detected observables relative to the prepared state
form only a semigroup into the forward light cone~\cite{schulman} expressing
Einstein causality.
\begin{equation}
P_+=\{(\Lambda,x):(\Lambda_0^0\ge1,\textrm{det}\Lambda=+1),x^2=t^2-\textbf{x}^2\ge 0, t\ge 0\}
\end{equation}

\newpage
\section{Resonances and Decay}
To find the theory that provides the Time Asymmetry semigroup, we use
scattering, resonance and decay phenomena \cite{physica}.
Resonances and decaying states are characterized by definite
values of the discrete quantum numbers such as charges (particles
species label) and by angular momentum $j$.
In addition they are defined by two real numbers.
Resonances are characterized by energy $E_R$ and width $\Gamma$,
and decaying states are characterized by energy $E_D$ and lifetime $\tau$.
Their properties are contrasted as follows:
\\[5pt]
\begin{minipage}{0.45\textwidth}
    $(E_{R},\Gamma)$ defined by
\end{minipage}\hfill
\begin{minipage}{0.45\textwidth}
    $(E_{D},R=\frac{1}{\tau})$ defined by\vspace*{5pt}
\end{minipage}\\
\begin{minipage}{0.45\textwidth}
    Breit-Wigner (Lorentzian) scattering amplitude.
\end{minipage}\hfill
\hspace*{10pt}\hfill\begin{minipage}{0.45\textwidth}
    exponential partial decay rate\vspace*{5pt}
\end{minipage}\\
\begin{minipage}{0.45\textwidth}
\vspace{10pt}
    $a_{j}^{BW}=\frac{r_{\eta}}{E-(E_{R}-i\frac{\Gamma}{2})}$; $0\leq E<\infty$
\end{minipage}\hfill
\begin{minipage}{0.45\textwidth}
    $R_{\eta}(t)=\frac{d}{dt}P_{\eta}(t)$, $R=\sum_{\eta}R_{\eta}(0)$\vspace*{5pt}
\end{minipage}\\
\hspace*{10pt}\hfill\begin{minipage}{0.45\textwidth}
    $R_{\eta}(t)=R_{\eta}(0)\text{e}^{-t/\tau}=R_{\eta}(0)\text{e}^{-Rt}$\vspace*{5pt}
\end{minipage}\\
\hspace*{10pt}\hfill\begin{minipage}{0.45\textwidth}
    where probability
    $P_{\eta}(t)=\vert\braket{\psi_{\eta}}{\phi_{D}(t)}\vert^{2}$\vspace*{5pt}
\end{minipage}\\
\\
\begin{minipage}{0.45\textwidth}
    Resonances appear in
\end{minipage}\hfill
\begin{minipage}{0.45\textwidth}
    Decaying states are
\end{minipage}\\
\begin{minipage}{0.45\textwidth}
    scattering, e.g.
\end{minipage}\hfill
\begin{minipage}{0.45\textwidth}
    observed in decay, e.g.
\end{minipage}\\
\begin{minipage}{0.45\textwidth}
\begin{center}
$\quad$\\
$\quad$\\
    $e^+\,e^-\rightarrow Z\rightarrow e^+\,e^-$
\end{center}
\end{minipage}\hfill

\begin{figure}[h!]
\begin{minipage}{0.45\textwidth}
\vskip-1cm
\label{quasistates} \setlength{\unitlength}{0.240900pt}
\begin{picture}(1000,150)(0,0)
\put(1300,150){\makebox(0,0){{$ \pi^{-} \, p \rightarrow \Lambda
\,K^{0}_{S} $}}} \put(1395,30){\rule{.400pt}{20pt}}
\put(1397,30){\vector(1,0){150}}
\put(1650,40){\makebox(0,0){{$\pi^+ \, \pi^- $ }}}
\end{picture}
\vskip-1cm
\end{minipage}
\end{figure}


\noindent
\begin{minipage}{0.45\textwidth}
    Resonances
\end{minipage}\hfill
\begin{minipage}{0.45\textwidth}
    Decaying States\vspace*{5pt}
\end{minipage}\\
\begin{minipage}{0.45\textwidth}
    are measured by Breit-Wigner lineshape in the cross section\\
\end{minipage}\hfill
\begin{minipage}{0.45\textwidth}
    are measured by the exponential law for the counting rate of the
    decay products $\eta$\vspace*{9pt}
\end{minipage}\\
\begin{minipage}{0.45\textwidth}
    $\sigma_{j}(E)\sim\vert a_{j}(E)\vert^{2}$\\ $=\left\vert
    \frac{r_{\eta}}{E-(E_{R}-i\frac{\Gamma}{2})} + B(E)
    \right\vert^{2}$\\
\\
 $B(E)$ is a slowly varying function of $E$ (background).
\\
\\
\end{minipage}\hfill
\begin{minipage}{0.45\textwidth}
    $R_{\eta}(t)\approx\frac{\Delta N(t_{i})}{\Delta
    t_{i}}\propto\text{e}^{-\frac{t}{\tau}}$
    \vskip 5pt
    where $\Delta N(t_{i})$ is the number of decay products
    registered in the detector during the time interval $\Delta
    t_{i}$ around $t_{i}$.\\
\end{minipage}
Examples:
\begin{figure}[h!]
\begin{minipage}{0.45\textwidth}
\includegraphics[width=9pc]{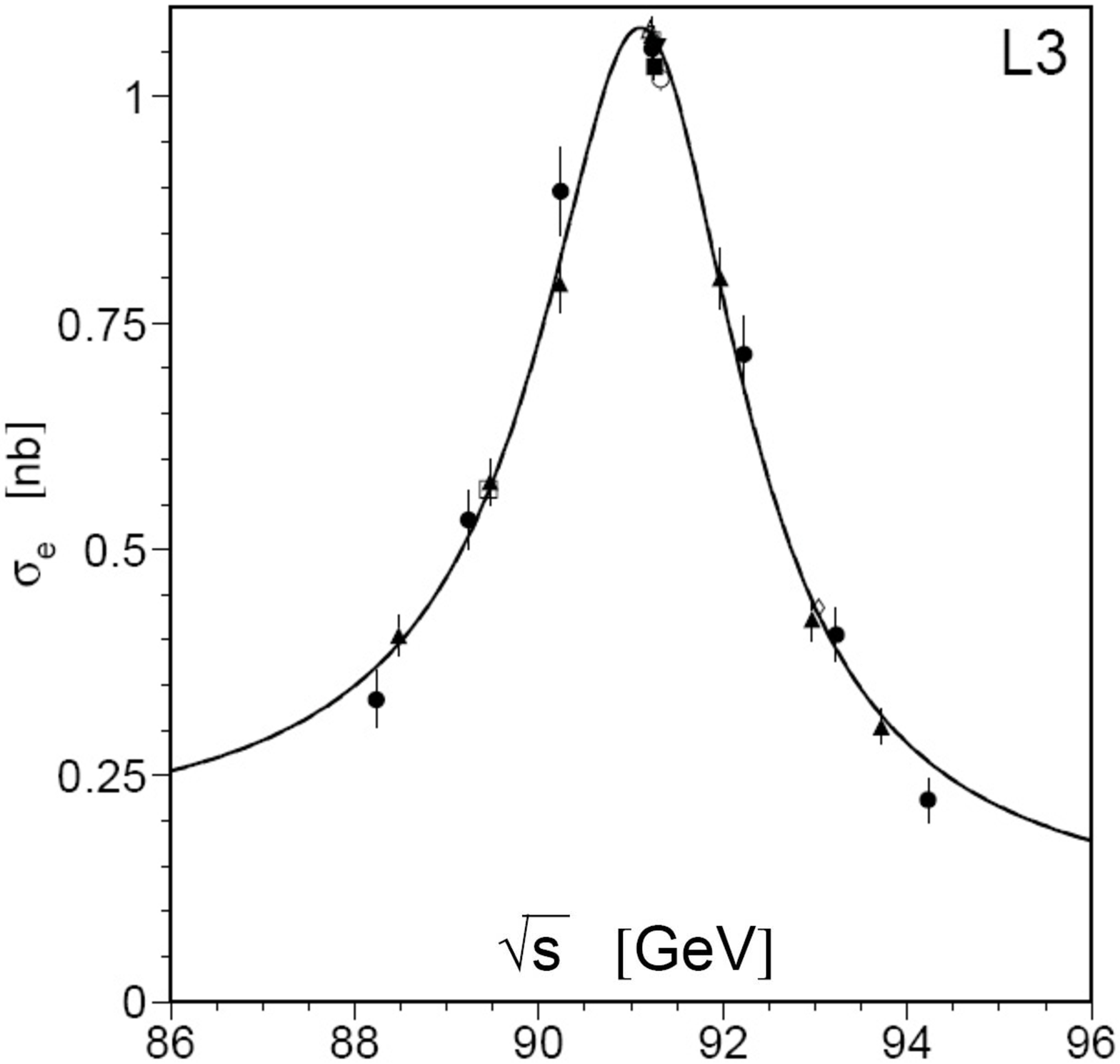}
\caption{\label{label1}Breit-Wigner for the $Z$-boson resonance \cite{fig1_ref}}
\end{minipage}\hfill
\begin{minipage}{0.45\textwidth}
$\quad$\\
\\
\includegraphics[width=12pc]{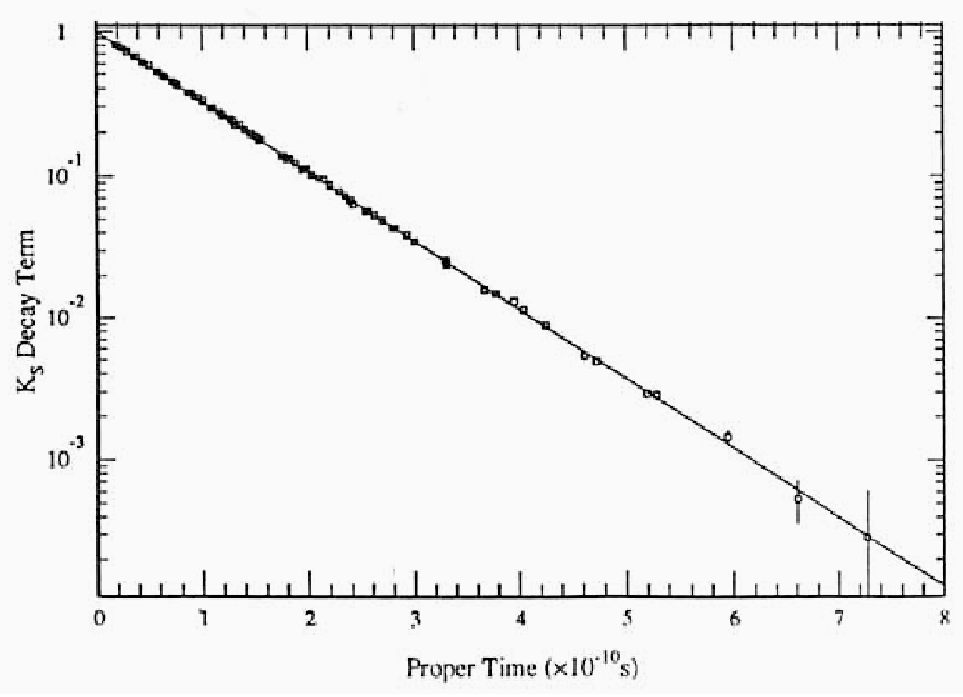}
\caption{\label{label2}Exponential for the $K_S^0$ decay rate \cite{gibbons}}
\end{minipage} 
\end{figure}

Many people think that
\begin{equation}
\{\text{resonances}\}\equiv\{\text{decaying states}\},
\end{equation}
and especially for non relativistic quantum mechanics, a common
assumption is that
\begin{equation}
\frac{\hbar}{\Gamma}=\tau\quad\left(\textrm{or at least }\frac{\hbar}{\Gamma}
\approx\tau\right).
\end{equation}
This relation is based on the Weisskopf-Wigner (WW) approximation
\cite{weisskopf_wigner}, and in
standard quantum theory there is no proof of it, as e.g. stated by 
\begin{quote}
    M.~Levy: \cite{levy} ``\ldots There does not exist\ldots\ a rigorous theory
    to which these various (WW) methods can be considered as
    approximations.''
\end{quote}
The energy of a resonance is the complex number\\[5pt]
\begin{minipage}{0.25\textwidth}
    $z_{R}=E_{R}-i\frac{\Gamma}{2}$,
\end{minipage}\hfill
\begin{minipage}{0.73\textwidth}
    the pole position of the $S$ matrix or of the scattering amplitude
    $a_{j}(E)$.
\end{minipage}\\
\\ \noindent
What one can show using the Weisskopf-Wigner approximation methods, is that the probability of a prepared
resonance state $\phi$ with Breit-Wigner energy distribution of width $\Gamma$ is
obtained as \cite{goldberger_watson}
\begin{equation}
\mathcal{P}_{\phi(t)}(\psi)\sim\text{e}^{-\Gamma
t/\hbar}+\Gamma\times(\text{additional terms}).
\end{equation}
Thus, in the Weisskopf-Wigner approximation, the probability rate has a non-exponential term
proportional to the width $\Gamma$.
In general one can prove that there is no Hilbert space vector $\phi(t)$
that obeys the exact exponential law \cite{khalfin}.
If, as in Figure \ref{label2} for the $K_S^0$-decay experiment, the time dependence of
$ \abs{\braket{\psi_{\pi\pi}}{\phi^{K^0_S}(t)}}^2=\frac{N(t_i)}{N}$ is to be
a perfect exponential (as shown by this experiment \cite{gibbons})
then the decaying state vector $\phi^{K^0_S}$ cannot be a Hilbert space vector.

\section{Complex Energies and a Beginning of Time}
The simplest way to derive an exactly exponential decay
probability is to postulate a state vector $\phi^{G}$, which has the
property
\begin{equation}
\label{honphi}
H\phi^{G}=(E_R-i\frac{\Gamma}{2})\phi^{G}\quad\text{and}\quad\phi^{G}(t)
=\text{e}^{-iHt}\phi^{G},
\end{equation}
and whose decay probability into any observable 
$\ketbra{\psi}{\psi}$ is
\begin{eqnarray}
\mathcal{P}_{\phi^{G}(t)}&=&\vert\braket{\psi}{\phi^{G}(t)}\vert^{2}
=\vert\langle\psi\mid\text{e}^{-iHt}\mid\phi^{G}(t)\rangle\vert^{2}\nonumber \\
\label{pofphi} &=&\vert\braket{\psi}{\phi^{G}}\text{e}^{-i(E_R-i\Gamma/2)t}\vert^{2}
=\vert\braket{\psi}{\phi^{G}}\vert^{2}\text{e}^{-\Gamma t}.
\end{eqnarray}
This vector $\phi^G$  of Gamow \cite{gamow} makes no sense in standard quantum
mechanics because:
\begin{enumerate}
  \item It has a complex eigenvalue for a selfadjoint $H$;\quad $\phi^G = \ket{E_R-i\Gamma/2}$
  \item It leads to the ``exponential catastrophe'' \cite{mainland}, because in standard
  quantum mechanics the time extends from $-\infty$: $-\infty<t<\infty$.
\end{enumerate}
To define a vector with the properties (\ref{honphi})--(\ref{pofphi}), one needs a theory for which
\begin{enumerate}
  \item the energy extends to values in the complex plane
  \item the time is restricted to $0\leq t<+\infty$ (because a $K^0_S$ must 
be prepared first before one can detect its decay products $\pi^+\,\pi^-$,
and to avoid the exponential catastrophe).
\end{enumerate}

In the standard (Hilbert space) theory of quantum
physics,
  the time $t$ has the values $-\infty<t<+\infty$,
  and the energy $E$ is real (spectrum of self adjoint
  Hamiltonian $H$) and bounded from below (stability of
  matter): $-\infty\ne E_{0}\leq E<\infty$.

Nevertheless, one speaks of complex energy:\\[5pt]
\hspace*{8pt}\begin{minipage}{0.6\textwidth}
    \begin{enumerate} 
\item[(i)]for the analytic $S$ matrix
    \end{enumerate}
\end{minipage}\hspace*{3pt}
\begin{minipage}{0.3\textwidth}
    $S_{j}(E)\rightarrow S_{j}(z)$
\end{minipage}\\[7pt]
\hspace*{8pt}\begin{minipage}{0.6\textwidth}
    \begin{enumerate}
\item[(ii)]for the Gamow states
    \end{enumerate}
\end{minipage}\hspace*{3pt}
\begin{minipage}{0.3\textwidth}
    $z_{R}=E_{R}-i\Gamma/2$
\end{minipage}\\[7pt]
\hspace*{8pt}\begin{minipage}{0.6\textwidth}
    \begin{enumerate}
\item[(iii)]for the Lippmann-Schwinger equation\\
or in the propagator of field theory
    \end{enumerate}
\end{minipage}\hspace*{3pt}
\begin{minipage}{0.4\textwidth}
    $z=E\pm i\epsilon$, $\epsilon$ infinitesimal
  \end{minipage}\\
\\
  Thus experiments require a quantum theory in which
\begin{enumerate}
  \item the time $t$ has a ``preferred direction'': $t_{0}=0\leq
  t<\infty$
  \item the energy $E$ 
  can take complex (discrete and continuous)
  values in the complex planes: $E\rightarrow z\in\mathbb{C_{\pm}}$.
\end{enumerate}
\begin{equation}
\ket{E} \longrightarrow \ \ket{E\pm i\epsilon}  \longrightarrow \ \ket{z^\pm} \quad z \in \mathbb{C}_\pm.
\end{equation}
The conclusion from this is that
one has to restrict the set of allowed energy wave functions
$\phi(E)=\braket{^+E}{\phi^+}$,
which in standard quantum theory in $\mathcal{H}$ are represented by a \textit{class} of Lebesgue square
integrable functions, to a smaller set.

The first step in this direction was already taken by Schwartz,
inspired by the Dirac bra-ket formalism,
when he restricted the set of \textit{classes} of Lebesgue square integrable functions
$\mathcal{L}^2=\{\{h(E)\}\}$ to the set of smooth, rapidly decreasing (Schwartz) functions
$\{\phi(E)\}$.  This replacement of the Hilbert space $\mathcal{H}$ by the Schwartz space
$\Phi$ gave a mathematical meaning to the Dirac formalism \cite{dirac}.

The second step is to restrict the set of Schwartz space functions $\phi(E)$ further to the set of smooth functions that have analytic extensions in the upper and lower complex energy semiplanes,
$\mathbb{C}_+$ and $\mathbb{C}_-$, respectively.
\begin{equation}
\label{changing_phi}
\{\overline{\phi(E)}=\braket{\phi}{E}\}\rightarrow
\begin{cases}
\braket{\phi^+}{E+i\epsilon}\equiv\braket{\phi^+}{E^+}\rightarrow\braket{\phi^+}{z^+};
&z\in \mathbb{C}_+\\
\braket{\psi^-}{E-i\epsilon}\equiv\braket{\psi^-}{E^-}\rightarrow\braket{\psi^-}{z^-};&z\in \mathbb{C}_-
\end{cases}
\end{equation}

\section{Rigged Hilbert Space}
The mathematical basis for these modifications (\ref{changing_phi}) are Rigged Hilbert Spaces
(RHS), or Gel'fand Triplets.
The RHS was not devised for Time Asymmetry or the theory of
resonances and decay, but to provide a mathematical justification for
Dirac's bra- and ket-formalism \cite{dirac}.

\begin{description}
\item[]Step 1 away from the Hilbert space was to get the Dirac formalism, which has the following
two properties:
\begin{enumerate}
\item
The solutions of both the Heisenberg and the Schr\"{o}dinger equations 
(for observables and states) have a Dirac basis vector expansion,
\begin{equation}
\label{vector_expansion}
\phi = \sum_{j,j_3}\sum_n |E_n,j,j_3)(E_n,j,j_3\ket{\phi}
	 + \sum_{j,j_3,\eta} \int dE\,\,\ket{E,j,j_3,\eta}\braket{E,j,j_3,\eta}{\phi},
\end{equation}
which is the analogue of $\vec{x}=\sum_{i=1}^3 \vec{e}_i x^i$,
and where the basis vectors $\ket{E,j,j_3,\eta}\equiv\ket{E}$ are ``eigenkets'' of $H$
(and of a complete system of commuting observables
for the other quantum numbers $j$, $j_3$, $\eta$).
The basis vectors in~\eqref{vector_expansion} are
defined as\footnote{
The first equality defines the conjugate operator $H^\times$ of $H$, $H^\times\supset H^\dagger$.}
the eigenkets of an operator $H$ 
\begin{subequations}
\begin{align}
\label{ket_def}
\braket{H\phi}{E,j,j_3,\eta}=\bra{\phi}H^\times\ket{E,j,j_3,\eta} &= E \braket{\phi}{E,j,j_3,\eta}\,\mbox{for all $\phi \in \Phi$}
,
\intertext{or as ordinary eigenvectors}
H\,|E_n,j,j_3,\eta) &= E_n\, |E_n, j, j_3 \eta)
\label{ket_def2}
.
\end{align}
\label{ket_def_all}
\end{subequations}
In (\ref{vector_expansion}) and (\ref{ket_def}),
the values of $E$ are from a continuous set, e.g. $0\leq E < \infty$.
\item
The co-ordinates or ``scalar products'' or the
bra-kets $\braket{E}{\phi}=\phi(E)$ are smooth, rapidly decreasing functions 
of $E$ (``Schwartz function'' $\mathcal{S}_\mathbb{R_+}$).
One gets
\begin{align}
\label{rhs1}
&\mbox{a Triplet of function spaces} \qquad \{ \phi(E) \} = \{\psi(E)\}=\mathcal{S}_\mathbb{R_+} \subset L^2 \subset \mathcal{S}^\times\\
&\mbox{and corresponding to this} \nonumber \\
\label{rhs2}
&\mbox{a Triplet of abstract vector spaces} \qquad \{ \phi \} = \{\psi \}=\Phi \subset \mathcal{H} \subset \Phi^\times \ni \ket{E}
\end{align}

The space $\Phi^\times$ denotes the space of continuous antilinear functionals of the space $\Phi$;
for the Hilbert space, $\mathcal{H}^\times = \mathcal{H}$.
Dirac kets are antilinear continuous Schwartz space functionals.
\end{enumerate}
The triplets (\ref{rhs1}) and (\ref{rhs2}) constitute  Rigged Hilbert Spaces.
($\Phi$ has a locally convex nuclear topology such that the 
Dirac basis vector expansion (\ref{vector_expansion}) is the nuclear spectral theorem.)
The RHS (\ref{rhs1}) of functions and distributions is equivalent (algebraically and topologically)
to the abstract RHS (\ref{rhs2}).

This Schwartz space triplet (of axiomatic quantum field theory) 
will not provide Time Asymmetry, because
solutions of the dynamical (Schr\"{o}dinger or Heisenberg) equation with boundary condition $\phi \in \Phi$, $\psi \in \Phi$ (Schwartz space) are again given by a group \cite{wick_02}
\begin{equation}
\phi(t) = e^{-iHt} \phi(0), \quad \psi(t) = e^{iHt} \psi(0), \quad -\infty< t < +\infty.
\end{equation}
For the Schwartz-Dirac kets one thus also obtains \cite{wick_02}
\begin{equation}
e^{-iH^\times t} \ket{E} = e^{-iEt} \ket{E} \quad -\infty < t < +\infty.
\end{equation}
This means Step 1 away from the Hilbert space provides the Dirac kets (\ref{ket_def}) and the Dirac basis
vector expansion (\ref{vector_expansion}),
but the Dirac formalism still does not lead to complex energy eigenvalues and resonance states (\ref{honphi}),(\ref{pofphi}).

\item[]Step 2:\\
Because observables $\ket{\psi^-}\bra{\psi^-}$ defined by the detector
and states $\phi^+$ defined by the preparation apparatus
are physically different entities, one should also distinguish 
mathematically between a \textit{space of observables}, which we call $\Phi_+$, 
\begin{equation}
\label{expansion1}
\Phi_+ \ni \psi^- = \sum_{j,j_3,\eta} \int_0^\infty dE\, \ket{E,j,j_3,\eta^-}\braket{^-E,j,j_3,\eta}{\psi^-}\\
\end{equation}
and a \textit{space of states}, which we call $\Phi_-$
\begin{equation}
\label{expansion2}
\Phi_- \ni \phi^+ = \sum_{j,j_3,\eta} \int_0^\infty dE\,
\ket{E,j,j_3,\eta^+}\braket{^+ E,j,j_3,\eta}{\phi^+}\, .
\end{equation}
The two Dirac basis vector expansions (\ref{expansion1}), (\ref{expansion2}) use 
different kinds of Dirac kets, which we denoted:
\begin{equation}
\ket{E,j,j_3,\eta^\mp}\in\Phi^\times_\pm.
\label{eq:ls_ket}
\end{equation}
This is suggested by the Lippmann-Schwinger in-plane waves $\ket{E^+}$
for the prepared in-states $\phi^+$, and the Lippmann-Schwinger out-plane
waves $\ket{E^-}$ for the detected out-observables $\psi^-.$\footnote{
The only major alteration of the Standard S-matrix formalism is that here we use
out-observables fulfilling the Heisenberg equation, and not out-states fulfilling the
Schr\"odinger equation.}

Because of the $+i\epsilon$ in the Lippmann-Schwinger equation,
the energy wave function of the prepared in-state $\phi^+$,
\begin{equation}
\phi^+(E) = \braket{^+E,j,j_3,\eta}{\phi^+} = \overline{\braket{\phi^+}{E,j,j_3,\eta^+}},
\end{equation}
is the boundary value of an analytic function in the lower complex 
energy semi-plane (for complex energy 
$z= \overline{E+i\epsilon}= E-i\epsilon$ immediately below the 
real axis 2nd sheet of the S-matrix $S_j(z)$).
Similarly, 
one surmises that 
the energy wave function of the observable
$\ketbra{\psi^-}{\psi^-},$
\begin{equation}
\psi^-(E)\equiv\braket{^- E,j,j_3,\eta}{\psi^-},
\end{equation}
extends to an analytic function in the upper complex energy plane.
Thus one conjectures that energy wave functions can only be those Schwartz functions that 
can be analytically continued into the lower or upper complex energy semiplane, respectively.
The space of allowed state vectors must be smaller than the Schwartz space, and the
space of allowed kets will be larger than the set of Dirac kets and thus can
include the eigenkets of complex energy.
\end{description}

\section{Conjecturing the Hardy Space Axiom of Time Asymmetric Quantum Mechanics}
To conjecture the spaces of $\phi^+$ and of $\psi^-$, we start with the definition of a resonance by an
S-matrix pole at $z_R=E_R-i\Gamma/2$.
If we derive from it
%
\vspace{-8pt}
\begin{align}
&\quad \mbox{a Breit-Wigner}\nonumber \\
&\mbox{resonance amplitude}\qquad \qquad \mbox{and relate to this Breit-Wigner a Gamow state vector} \nonumber\\
&\qquad a_j^{BW_i}=\frac{R_i}{E-z_{R_i}} \quad  \Longleftrightarrow \quad \phi_j^G = \ket{z_{R_i},j,j_3,\eta^-}\sqrt{2\pi \Gamma_i}=\int dE\, \ket{E,j,j_3,\eta^-}\frac{i\sqrt{\frac{\Gamma}{2\pi}}}{E-z_R}
\label{eq:bw_amp}\\
&\quad \mbox{for every pole $z_{R_i}$}\qquad\qquad\qquad\qquad\qquad \mbox{$z_{R_i}= E_{R_i}-i\Gamma_i/2$}
\nonumber 
\end{align}
which has the properties:\\
\textbf{1.} it is an eigenket with a discrete complex eigenvalue (as Gamow wanted) of the Hamiltonian:
\begin{align}
&H^\times \ket{E_R-i\Gamma/2^-}= (E_R-i\Gamma/2)\,\ket{E_R-i\Gamma/2^-},
\quad\ket{E_R-i\Gamma/2^-}\in\Phi^\times_+,
\label{eq:gamow_eigenvalue}
\end{align}
and \textbf{2.} it has the time evolution
\begin{eqnarray}
\braket{\mathcal{U}(t)\psi_\eta^-}{\phi_j^G}
&=&\braket{\psi_\eta^-}{\mathcal{U}^\times(t)\phi^G_j}\sim\nonumber\\
\braket{e^{iHt/\hbar}\psi_\eta^-}{E_R-i\Gamma/2^-} 
&=&\bra{\psi_\eta^-}e^{-iH^\times t/\hbar}\ket{E_R-i\Gamma/2^-}\nonumber\\
&=&e^{-iE_R t/\hbar} e^{-(\Gamma/2)t/\hbar} 
\braket{\psi_\eta^-}{E_R-i\Gamma/2^-}\quad\textrm{for all }\psi_\eta^-\in\Phi_+,
\label{eq:gamow_time}
\end{eqnarray}
then we will have shown that:
\begin{equation}
\textrm{A resonance of width }\Gamma\equiv
\textrm{a decaying state with lifetime } \tau=\frac{\hbar}{\Gamma},
\end{equation}
and we will have a theory that unites resonance scattering and exponential decay.\\

To derive these results, further conditions had to be put on the 
analytic wave functions $\phi^+(z)$ and $\overline{\psi^-(z)}$ in the lower complex
semiplane second sheet.
These conditions suggested Hardy functions (H. Baumgartel).

Thus one is led to a \textbf{new hypothesis} \cite{physica}:\\
The energy wave functions of a state are smooth Hardy functions
analytic on the lower complex semiplane $\mathbb{C}_-$:
\begin{equation}
\phi^+(E) = \braket{^+E}{\phi^+} \in (\mathcal{H}^2_-\cap \mathcal{S})_{\mathbb{R_+}}.
\end{equation}
The energy wave functions of an observable are smooth Hardy functions
analytic on the upper complex semiplane $\mathbb{C}_+$:
\begin{equation}
\psi^-(E) = \braket{^-E}{\psi^-} \in (\mathcal{H}^2_+\cap \mathcal{S})_{\mathbb{R_+}}.
\end{equation}

This hypothesis is not so far off from the properties that one often
used for mathematical manipulations in scattering theory.
This new hypothesis led to the construction of the two
Hardy space triplets \cite{gadella}, i.e. two RHS's from which the two Dirac basis vector
expansions (\ref{expansion1}),(\ref{expansion2}) follow as the nuclear spectral theorem
\cite{gelfand}.

Therewith we have arrived at a new axiom of quantum mechanics, which is
the {\it Hardy space axiom}:\\[10pt]
\begin{minipage}{0.45\textwidth}
The set of prepared (in-) states defined by the
preparation apparatus (e.g. accelerator)
\end{minipage}\hfill
\begin{minipage}{0.45\textwidth}\begin{center}
is represented by
\begin{equation}
\{\phi^+\}=\Phi_- \subset \mathcal{H} \subset \Phi_-^\times
\label{eq:state_hrhs}
\end{equation}\end{center}
\end{minipage}\\[15pt]
\begin{minipage}{0.45\textwidth}
The set of (out-) observables defined by the
registration apparatus (e.g. detector)
\end{minipage}\hfill
\begin{minipage}{0.45\textwidth}\begin{center}
is represented by
\begin{equation}
\{\psi^-\}=\Phi_+\subset \mathcal{H} \subset \Phi_+^\times
\label{eq:obs_hrhs}
\end{equation}\end{center}
\end{minipage}\\[10pt]
\\
\noindent
Here $\Phi_\mp$ are Hardy spaces of the semiplanes $\mathbb{C}_\mp$.
The kets $\ket{z^\pm}$,\ $\ket{E^\pm}$,\ $\ket{E_R-i\Gamma/2^\pm}$ are 
functionals on the space $\Phi_\mp$.
In particular, the exponentially decaying
Gamow ket $\ket{E_R-i\Gamma/2^-}$ is a functional on the space $\Phi_+$.

Experimentalists distinguish
between preparation apparatus (accelerator) 
and registration apparatus (detector).
In the foundations of quantum mechanics, one talks of states and observables
as different entities \cite{kraus,haag}.
In Hilbert space theory one identifies them mathematically.
The Hardy space  axiom distinguishes also mathematically
between states, $\phi^+\in\Phi_-$,
and observables, $\psi^-\in\Phi_+$, as different dense subspaces
of the same Hilbert space $\mathcal{H}$.
This is entirely natural.

\section{Semigroup Time Evolution}
The dynamical equations of quantum mechanics (\ref{heis_eqn}) or (\ref{schrod_eqn}) can now be solved for the
observable $\psi^-$ and the state $\phi^+$ under the Hardy space boundary condition
$\psi^-\in\Phi_+$ and $\phi^+\in\Phi_-$.
As a consequence of the Paley-Wiener theorem \cite{koosis} it follows:
\begin{center}
The solutions of dynamical equations\\[5pt]
\noindent
\begin{tabular}{ccc}
(Heisenberg Equation):  & \quad & (Schr\"{o}dinger Equation):\\
of observables in space $\Phi_+$ & and & of states in the space $\Phi_-$\\
\quad & \quad & \quad \\
\quad & are given by the semigroups & \quad \\
\end{tabular}
\end{center}
\begin{minipage}{0.4\textwidth}
\begin{eqnarray}
\psi^-(t)=e^{iHt} \psi^-\\
t_0=0 \le t < \infty \nonumber 
\end{eqnarray}
\end{minipage}\hfill
\begin{minipage}{0.4\textwidth}
\begin{eqnarray}
\phi^+(t) = e^{-iHt} \phi^+\\
t_0=0 \le t < \infty\nonumber
\end{eqnarray}
\end{minipage}\\[10pt]
\noindent
The Lippmann-Schwinger scattering states in $\Phi^\times_+$ fulfill:
\begin{equation}
e^{-iH^\times t} \ket{Ej\eta^-}=e^{-iEt} \ket{Ej\eta^-}\qquad 0 \le t < \infty.
\end{equation}
Therefore the probability for the time evolved observable 
$\psi^-(t)$ in the state $\phi^+$ can now be calculated:
\begin{equation}
\mathcal{P}_{\phi^+}(\psi^-(t)) 
= \abs{\braket{\psi^-(t)}{\phi^+}}^2
= \abs{\braket{e^{iHt}\psi^-}{\phi^+}}^2
= \abs{\braket{\psi^-}{e^{-iH^\times t} \phi^+}}^2\quad
\textrm{for }t\ge t_0=0\textrm{ only}.
\end{equation}
This is in agreement with the phenomenological conclusion at which we arrived in
(\ref{born_arrow}) on the basis of causality.
For the detector counts,
this means that
\begin{equation}
\mathcal{P}_{\phi^+}(\psi^-(t)) \sim N(t)/N \quad
\mbox{can be measured only for $t\geq 0$},
\end{equation}
i.e. after the state has been prepared.

As a special case, the probability for the decay products
$\ketbra{\psi_\eta^-}{\psi_\eta^-}$
in a Gamow state $\phi^G=\ket{E_R-i \Gamma /2^-}$
is predicted to be
\begin{equation}
\vert \braket{e^{i H t/\hbar}\psi^-_\eta}{\phi^G} \vert^2=e^{-\Gamma t/\hbar }
\vert \braket{\psi^-_\eta}{\phi^G} \vert^2 \quad \textrm{only for } t\geq 0.
\label{eq:gamow_prob}
\end{equation}
This avoids the ``exponential catastrophe'' \cite{mainland} for Gamow states.

\section{Application: Correct values of mass and width of the $Z$-boson and other relativistic resonances}
\label{application_section}
	
	Causal evolution of the non-relativistic spacetime can be extended to relativistic spacetime. 
	Whereas non-relativistic time evolution is described by the Galilei group with invariants $m$, $E$, $j$, $\eta$ (particle species or channel number), the relativistic time evolution is described by the Poincar\`e group $\mathcal{P}$ with invariants $\s=p^\mu p_\mu$ and $j$, spin, for a particle species $\eta$ \cite{weinberg}.
	The causal time evolution is then given by the Poincar\`e semigroup into the forward light cone,

\begin{align}
\mathcal{P}_+ = \left\{ (\Lambda,x) : x^2 = t^2 - \bm{x}^2 \geq 0, t\geq 0 \right\}
.
\end{align}
	In analogy to Wigner's unitary group representations $[\s=m^2,j]$ for stable particles, the relativistic resonance particles are described by semigroup representations into the forward light cone with invariants $[\s_R, j]$~\cite{ref:PRD}, where $\s_R$ is a complex mass squared given by the pole position of the relativistic S-matrix.
	This means that the resonance amplitude $a_j^{\rm res}(\s)$ is given by
\begin{align}
a_j^{\rm res}(\s) = \frac{r}{\s-\s_R}
.
\end{align}
	The basis vectors of the semigroup representation $[\s_R,j]$ are analogous to the Wigner kets given by  the relativistic Gamow ket $\ket{[\s_R,j]\vel{p}j_3\,^-}$~\cite{ref:PRD}.

	With only the complex number for the pole position $\s_R$ given, there can be many parameterizations in terms of two real parameters, $(M, \Gamma)$.
	For instance, for the $Z$-boson one used
\begin{subequations}
\label{overall}
\begin{align}
\s_R &= (M_R-i\Gamma_R/2)^2
\label{eq:correctone}
\\
\s_R &= \overline{M}_Z^2-i\overline{M}_Z\overline{\Gamma}_Z
.
\end{align}
\label{eq:smatrixpoles}
\end{subequations}
	In addition, a very popular parameterization is the on-the-mass-shell definition of $(M_Z, \Gamma_Z)$, obtained from the propagator definition in the on-mass-shell
renormalization scheme~\cite{pdg_book}:
\begin{align}
a_j^{\rm res}(\s) = \frac{R_Z}{\s-M_Z^2+i\frac{\s}{M_Z}\Gamma_Z}
\label{eq:onmassshell}
.
\end{align}
	The values for these different parameterizations $(M,\Gamma)$ are obtained from fits of $a_j(\s)=a_j^{\rm res}(\s)+B_j(\s)$ to the cross section data (``line shape'') $\abs{a_j(\s)}^2$.  ($B(\s)$ is a slowly varying background.)
	Depending upon the different choices \eqref{overall} and \eqref{eq:onmassshell}, this leads to the different ``experimental'' values of the resonance mass ($M_Z$ and $\overline{M}_Z$ are listed in PDG book) as given in Table~\ref{tab:zmasswidth}:
\begin{table}[h]
\caption[$Z$-boson masses and widths.]{$Z$-boson masses and widths from PDG}
\begin{center}
\begin{tabular}{ccc}
\br
$M_Z=91.1875\pm0.0021 {\rm GeV}$ & $M_R=91.1611\pm0.0023{\rm GeV}$ & $\overline{M}_Z=91.1526\pm0.0023{\rm GeV}$ \\
$\Gamma_Z=2.4939\pm0.0024{\rm GeV}$ & $\Gamma_R=2.4943\pm0.0024{\rm GeV}$ & $\overline{\Gamma}_Z=2.4945\pm0.0024{\rm GeV}$ \\
\br
\end{tabular}
\label{tab:zmasswidth}
\end{center}
\end{table}
	The situation for the other well measured (hadron) resonances $\Delta$ and $\rho$ is similar.
	Therefore the question is: Which is the correct pair of mass and width?
	Or is there no right value---is the value of $M$ and $\Gamma$ just a matter of convention for the parameterization of the complex value $\s_R$?

	Using, in analogy to Wigner's definition for stable relativistic particles, the definition
by causal relativistic spacetime transformations as the definition for the mass of a relativistic resonance, one can fix the values of $M$ and $\Gamma$ in the parameterization~\eqref{eq:smatrixpoles} uniquely and exclude the parameterization~\eqref{eq:onmassshell}.
From the time evolution of a relativistic Gamow vector of the semigroup representation $[\s_R,j]$, one calculates (for simplicity here in the rest frame $\vel{p}=\bm{0}$)
\begin{align}
H^\times\,\ket{[\s_R,j]\vel{p}=\bm{0}\,j_3\,^-} = \sqrt{\s_R}\,\ket{[\s_R,j]\vel{p}=\bm{0}\,j_3\,^-}
,
\end{align}
where $H^\times = P_0$.
	If we take the parameterization~\eqref{eq:correctone}, the action of a time translation
on the relativistic Gamow vector is given by
\begin{align}
\phi_{\s_R}^G(t) 
&= e^{-iH^\times t/\hbar}\, \ket{[\s_R,j]\vel{p}=\bm{0}\,j_3\,^-}
= e^{-iM_R\,t/\hbar}e^{-(\Gamma_R/2)t/\hbar}\,\ket{[\s_R,j]\vel{p}=\bm{0}\,j_3\,^-} 
\label{eq:timeevolvingrgv}
.
\end{align}
	The Born probability density for detecting the decay products (observable $\ketbra{\psi^-}{\psi^-}$) in the quasistable state $\phi_{\s_R,j}^G(t) = \ket{Z^-}$ at time $t\geq 0$ is then, as a consequence of \eqref{eq:timeevolvingrgv}, proportional to
\begin{align}
\abs{\braket{\psi^-}{\phi_{\s_R}^G(t)}}^2 = e^{-\Gamma_R t}\abs{\braket{\psi^-}{\phi_{\s_R}^G(0)}}^2
. 
\end{align}

	From this we conclude that the Gamow vector with the relativistic Breit-Wigner line shape $1/(\s-\s_R)$ and the parameterization of the S-matrix pole position given by  $\s_R=(M_R-\Gamma_R/2)^2$ has the lifetime $\tau=\hbar/\Gamma_R$.
	Therefore $(M_R, \Gamma_R)$ is the correct definition of $(M, \Gamma)$ for a relativistic resonance.
	With this, the ``right'' values of mass and width of the $Z$-boson are
\begin{align}
M_R &=\quad{\rm Re}\sqrt{\s_R} = 91.1611\pm0.0023\,{\rm GeV} = M_Z-0.0026\,{\rm GeV}
\\
\Gamma_R &= -2{\rm Im}\sqrt{\s_R} = 2.4943\pm0.0024\,{\rm GeV}
.
\end{align}

\section{Conclusion}
The Hardy space axiom is a refinement of the standard axiom of quantum mechanics.
Standard Hilbert space quantum mechanics works fine for spectra and structure of
micro-physical systems in stationary states.
For dynamically evolving states, for resonance scattering 
and for decaying states, the
Hardy space axiom works better and provides a theory that
unifies resonance and decay phenomena.
If one admits more general operators for observables and states
(as mentioned in the Epilogue)
one also obtains exponentially decaying states that are associated
with S-matrix poles of order $\mathcal{N}>1$, but these states
cannot be described by vectors or kets.
They are described by non-diagonalizable operators \eqref{eq:density_op}, \eqref{eq:density_def},
and their Hamiltonians contain non-diagonalizable Jordan matrices.
The Hardy space axiom
also introduces a novel concept: a quantum mechanical beginning of time,
or the semigroup time $t_0=0$.
This concept is not entirely new.
In Gell-Mann and Hartle's quantum theory of the universe \cite{gman},
this $t_0=t_{\textrm{big bang}}$.
But how does one see this time $t_0=0$ in the usual experiments with quantum
systems in the lab?
The discussion of this question needs to be postponed to another paper.

\section*{Epilogue}
\appendix
\setcounter{section}{1}
	In order to be as intelligible as possible, the talk at QTS-5 was confined to the 
    simplest cases possible: the observables were represented by vectors $\psi^-$, 
    i.e., by $\Lambda=\ketbra{\psi^-}{\psi^-}$; the states were represented by 
    vectors $\phi^+$, and for the decaying states we used the Gamow state vector $\phi^G$ of~\eqref{eq:gamow_eigenvalue}.
	However observables are in general represented not by $\ket{\psi^-}$ but by operators 
    that obey the Heisenberg equation~\eqref{heis_eqn}, and states are in general described by 
    density operators $W$ obeying the von-Neumann equation~\eqref{eq:density_eom}.
	Some of these generalizations are accommodated in a straightforward way, e.g., the observable one generalizes $\ketbra{\psi^-}{\psi^-} \longrightarrow \Lambda = \int dE\,\lambda(E)\,\ketbra{^-E}{E^-}$.
	Or for the state one goes to density operators $\ketbra{\phi^+}{\phi^+} \longrightarrow W$ so that the Born probabilities~\eqref{eq:born_prob1} become  $\abs{\braket{\psi^-}{\phi^+(t)}}^2 \longrightarrow {\rm Tr}(\Lambda W(t))$.
	This can be done when the Hamiltonian $H$ can be diagonalized, as is always the case for selfadjoint $H$ in the Schwartz space with the basis system~\eqref{ket_def},~\eqref{ket_def2} in~\eqref{vector_expansion}.
%
%
	The Dirac basis vector expansion in the matrix representation \eqref{ket_def_all} is given as
\begin{align}
\left(
  \begin{array}{c}
    \bra{\psi}H^\times|E_1) \\
    \bra{\psi}H^\times|E_2) \\
    \vdots \\
    \bra{\psi}H^\times|E_N)\\
    \bra{\psi}H^\times\ket{E^-}
  \end{array}
  \right)
=
\left(
  \begin{array}{ccccc}
    E_1 & 0   & \cdots & \cdots & 0 \\
    0   & E_2 & \quad  & \quad  & 0 \\
    \vdots & \quad & \ddots & \quad & \vdots \\
    \vdots & \quad & \quad & E_n & 0 \\
    0 & 0 & \cdots & 0 & (E)
  \end{array}
  \right)
\left(
  \begin{array}{c}
    \bra{\psi}E_1) \\
    \bra{\psi}E_2) \\
    \vdots \\
    \bra{\psi}E_N)\\
    \braket{\psi}{E^-}
  \end{array}
  \right)
\label{eq:matrix1}
\end{align}
where $(E)$ denotes a continuously infinite submatrix, and $E$ takes the values $0\leq E<+\infty$ in the diagonal
with the off-diagonal elements being zero.

	If one uses instead of the Hilbert space axiom~\eqref{eq:hilbert_bc} with 
    Dirac expansion~\eqref{vector_expansion} the Hardy space axiom~\eqref{eq:state_hrhs},~\eqref{eq:obs_hrhs}, 
    the energy wave functions $\braket{\psi^-}{E^-}\in \mathcal{H} \cap \mathcal{S}$ can be 
    continued into the lower complex energy plane (2nd sheet), and Gamow 
    vectors~\eqref{eq:gamow_eigenvalue} -- generalized eigenvectors with complex generalized 
    eigenvalue $z_R=E_R-i\Gamma/2$ -- can appear.
	The Gamow vector~\eqref{eq:gamow_eigenvalue} is associated with a first order pole of the 
    S-matrix (2nd sheet) at $z_R$~\cite{physica}.
	Assume that there are two resonances, at $z_{R_1}$ and $z_{R_2}$, and assume there are no discrete 
    energy eigenvectors (no bound state) with $E_1, E_2, \cdots, E_N$ in~\eqref{vector_expansion}.  
    Then the matrix representation of the selfadjoint Hamiltonian has the following form:
\begin{align}
\left(
  \begin{array}{c}
    \braket{H\psi^-}{z_{R_1}^-} \\
    \braket{H\psi^-}{z_{R_2}^-} \\
    \braket{H\psi^-}{E^-}
  \end{array}
\right)
=
\left(
  \begin{array}{c}
    \bra{\psi^-}H^\times\ket{z_{R_1}^-} \\
    \bra{\psi^-}H^\times\ket{z_{R_2}^-} \\
    \bra{\psi}H^\times\ket{E^-}
  \end{array}
  \right)
=
\left(
  \begin{array}{ccc}
    z_{R_1} & 0       & 0 \\
    0       & z_{R_2} & 0 \\
    0       & 0       & (E)
  \end{array}
  \right)
\left(
  \begin{array}{c}
    \braket{\psi^-}{z_{R_1}^-} \\
    \braket{\psi^-}{z_{R_2}^-} \\
    \braket{\psi^-}{E^-}
  \end{array}
  \right)
\label{eq:matrix2}
\end{align}
where $(E)$ denotes again a continuously infinite submatrix with real values in the diagonal.
	This -- with one resonance at $z_R$ -- is the case we have restricted ourselves to in the main part of the paper.
	But the mathematical theory we have devised for a Breit-Wigner resonance provides much more. 


	Because the Hardy space $\Phi_+$ is contained in the Schwartz space $\Phi$,
    its dual $\Phi_+^\times$ is much richer than $\Phi^\times$:
	In addition to the Lippmann-Schwinger kets~\eqref{eq:ls_ket} with real energy $E$ and their analytic continuations $\ket{z^-}$ -- which are tacitly assumed in any analytic S-matrix theory -- and in addition to the ordinary Gamow kets~\eqref{eq:gamow_eigenvalue}, the space $\Phi_+^\times$ also contains $\mathcal{N}$ dimensional subspaces $\mathcal{M}_{z_\mathcal{N}}\subset \Phi_+^\times$ where $\mathcal{N}$ can be $\mathcal{N}=1,2,\cdots$ any finite number.
	The subspace $\mathcal{M}_{z_\mathcal{N}}$ is spanned by $\mathcal{N}$ Jordan vectors with complex generalized eigenvalue $z_\mathcal{N}=\left(E_\mathcal{N}-i\Gamma_\mathcal{N}/2\right)$,
\begin{align}
|z_\mathcal{N}^-\succ^{(0)},\,|z_\mathcal{N}^-\succ^{(1)},\cdots, |z_\mathcal{N}^-\succ^{(k)},\cdots, |z_\mathcal{N}^-\succ^{(\mathcal{N}-1)}.
\label{eq:jg_vector}
\end{align}
%
($\mathcal{M}_{z_{\mathcal{N}=1}}$ is the space spanned by the ordinary 
Gamow ket in~\eqref{eq:gamow_eigenvalue} $\ket{z_1^-}^{(0)} =\,\ket{E_R-i\Gamma/2^-}$.)
	The $k$-th order Gamow vector $|z_\mathcal{N}^-\succ^{(k)}$, $k=0,1,\cdots,(\mathcal{N}-1)$, 
    is a Jordan vector of degree $(k+1)$, i.e., it fulfills the generalized eigenvalue 
    equations~\cite{ref:baumgaertel84, ref:bohm97}   
\begin{align}
&(H^\times-z_\mathcal{N})^{k+1}|z_\mathcal{N}^-\succ^{(k)} = 0;
\nonumber
\\
&H^\times|z_\mathcal{N}^-\succ^{(0)} = z_\mathcal{N} |z_\mathcal{N}^-\succ^{(0)} ;
\nonumber
\\
&H^\times|z_\mathcal{N}^-\succ^{(k)} = z_\mathcal{N}|z_\mathcal{N}^-\succ^{(k)} + \Gamma_\mathcal{N} |z_\mathcal{N}^-\succ^{(k-1)}\,
\mbox{for $k=1,2,\cdots, (\mathcal{N}-1)$}
\label{eq:jg_eigenvalue}
\end{align}
	These equations are, like the eigenvector equation for Dirac kets~\eqref{ket_def} and for Gamow vectors~\eqref{eq:gamow_eigenvalue} (Gamow vector = Jordan vectors of degree 1), understood as generalized eigenvector equations (i.e., functionals) over the space $\Phi_+$.
	Jordan block matrices for non-Hermitian Hamiltonians have been discussed before, 
    e.g.~\cite{ref:horwitz69,ref:katznelson80}, and were used for finite dimensional phenomenological 
    expressions of the S-matrix~\cite{ref:mondragon94, ref:hernandez93, ref:mondragon95, ref:hernandez92}, 
    but could not be implemented in the general framework of quantum mechanics using the Hilbert space or the Schwartz space axiom.

	Using the Hardy space axiom~\eqref{eq:state_hrhs}~\eqref{eq:obs_hrhs} the Jordan-Gamow vectors can be derived from the $\mathcal{N}$-th order pole of the unitary S-matrix~\cite{ref:antoniou98, ref:bohm97} in very much the same way as the ordinary Gamow vectors were derived from the first order S-matrix pole~\eqref{eq:bw_amp}~\cite{physica}.
	The matrix of the Hamiltonian $H^\times$ has in the diagonal the complex eigenvalues $z_R=E_R-i\Gamma_R/2$ of the 1st order pole position for the ordinary Gamow kets. 
	In addition it contains Jordan blocks of Jordan-Gamow kets.
	For example in the case of two first order poles at $z_{R_1}$ and $z_{R_2}$ and one second order pole at $z_2=E_2-i\Gamma_2/2$ the matrix of the Hamiltonian is given by
\begin{align}
\left(
  \begin{array}{c}
    \bra{H\psi^-}z_2^-\succ^{(0)} \\
    \bra{H\psi^-}z_2^-\succ^{(1)} \\
    \braket{H\psi^-}{z_{R_1}} \\
    \braket{H\psi^-}{z_{R_2}} \\
    \braket{H\psi^-}{E}
  \end{array}
\right)
=
\left(
  \begin{array}{c}
    \bra{\psi^-}H^\times |z_2^-\succ^{(0)} \\  
    \bra{\psi^-}H^\times |z_2^-\succ^{(1)} \\  
    \bra{\psi^-}H^\times\ket{z_{R_1}} \\
    \bra{\psi^-}H^\times\ket{z_{R_2}} \\
    \bra{\psi^-}H^\times\ket{E}
  \end{array}
\right)
=
\left(
  \begin{array}{ccccc}
	z_2 & 0 & \quad & \quad & \quad\\
	\Gamma_2 & z_2 & \quad & \quad & \quad\\
	\quad & \quad & z_{R_1} & \quad & \quad\\
    \quad & \quad & \quad & z_{R_2} & \quad\\
    \quad & \quad & \quad & \quad & (E)
  \end{array}
  \right)
\left(
  \begin{array}{c}
    \bra{\psi^-}z_2^-\succ^{(0)} \\  
    \bra{\psi^-}z_2^-\succ^{(1)} \\  
    \braket{\psi^-}{z_{R_1}} \\
    \braket{\psi^-}{z_{R_2}} \\
    \braket{\psi^-}{E}
  \end{array}
\right)
\label{eq:matrix3}
\end{align}
	Here $(E)$ denotes the continuously infinite matrix with diagonal elements $E$, as in~\eqref{eq:matrix2}.
	Each $z_{R_i}$ corresponds to the ordinary Gamow ket~\eqref{eq:gamow_eigenvalue}.
	And the $2\times2$ matrix in the top left corner is the Jordan block~\eqref{eq:jg_eigenvalue} of degree 2.

	In general, from the $\mathcal{N}$-th order S-matrix pole at $z_\mathcal{N}$ one obtains the $\mathcal{N}$ basis vectors~\eqref{eq:jg_vector} and an $\mathcal{N}$-dimensional Jordan block~\eqref{eq:jg_eigenvalue} in place of the two dimensional Jordan block in the matrix~\eqref{eq:matrix3}.
	This means the 2nd order or $\mathcal{N}$-th order pole of the S-matrix can no longer be described by a state vector, like the bound states by $|E_n)$ with real discrete eigenvalues, or the 1st order resonance states (Gamow states) by $\ket{z_{R_i}^-}$ with complex eigenvalue $z_{R_i}$ of the selfadjoint Hamiltonian $H$.
	Instead, the state derived from the $\mathcal{N}$-th order S-matrix pole is described by a non-diagonalizable density operator or state operator~\cite{ref:bohm97}
\begin{align}
W_{PT} = 2\pi\Gamma \sum_{n=0}^{\mathcal{N}-1} \left( \begin{array}{c}\mathcal{N}\\n+1\end{array}\right) (-i)^n W^{(n)}
\label{eq:density_op}
\end{align}
where the operators $W^{(n)}$ are defined as
\begin{align}
W^{(n)} = \sum_{k=0}^n |z_\mathcal{N}^-\succ^{(k)}\,^{(n-k)}\prec^-z_\mathcal{N}|
,\quad n=0,1,2,\cdots,\mathcal{N}-1
\label{eq:density_def}
.
\end{align}
	The pole term of the $\mathcal{N}$-th order S-matrix is associated with a sum~\eqref{eq:density_op} of the operators $W^{(n)}$.
	The operators $W^{(n)}$ represent components of this pole term state $W_{PT}$ which are ``irreducible'' in a way specified below in~\eqref{eq:wn_time}.
	In the case $\mathcal{N}=1$ (ordinary first order resonance pole) the operator~\eqref{eq:density_op} becomes
\begin{align}
W_{PT} = 2\pi\Gamma |z_1^-\succ^{(0)}\,^{(0)}\prec^-z_1| = 2\pi\Gamma W^{(0)} = \ketbra{\phi^G}{\phi^G}
.
\end{align}
	This is the operator description of the Gamow state whose vector description is given by $\phi^G$ of~\eqref{eq:gamow_eigenvalue} and whose time evolution is, in agreement with~\eqref{eq:gamow_time}, given by 
\begin{align}
W^G(t) 
&= e^{-i H^\times t} \ketbra{\phi^G}{\phi^G} e^{iHt}
= e^{-i z_R t} \ketbra{\phi^G}{\phi^G} e^{iz_R^*t}
\nonumber \\
&= e^{-i (E_R-i\Gamma/2) t} \ketbra{\phi^G}{\phi^G} e^{i(E_R+i\Gamma/2)t}
= e^{-\Gamma t} W^G(0)
\label{eq:wg_time}
.
\end{align}

	Only the 0-th order Gamow vector $|z^-\succ^{(0)}=\frac{1}{\sqrt{2\pi\Gamma}}\phi^G$ has exponential time evolution.
	In general, the kets $|z_\mathcal{N}^-\succ^{(k)}$, $k=0,\cdots,\mathcal{N}-1$ have very complicated time evolution given by
\begin{align}
e^{-iH^\times t}|z_\mathcal{N}^-\succ^{(k)} 
=
e^{-iz_\mathcal{N} t}\sum_{\nu=0}^k \frac{\Gamma^\nu}{\nu!} (-it)^\nu |z_\mathcal{N}^-\succ^{(k-\nu)} \quad t\geq 0
\label{eq:general_time}
.
\end{align}
	These are representations of the time translation semigroup, which (for $\mathcal{N}>1$) are not one dimensional.
	The existence of this kind of representation for the causal spacetime translation group has already been mentioned in~\cite{schulman}.
	The appearance of a linear term in the time dependence for a 2nd order pole resonance 
    $\mathcal{N}=1$ in~\eqref{eq:general_time} has been well known for many 
    years~\cite{goldberger_watson, ref:newton66, ref:goldberger64}.
	That such linear time-dependence has never been observed was used as an argument against the existence of higher order pole resonances~\cite{ref:goldberger64}.
    This was a misconception because the state associated to the S-matrix pole is described
    by a state operator (density matrix) \eqref{eq:density_op}, which has an exponential time 
    evolution, as we shall report now.

	The density operator or statistical operator of the state \textit{derived} 
    from the $\mathcal{N}$-th order pole is given by~\eqref{eq:density_op}~with~\eqref{eq:density_def}.
	It is remarkable that with the use of~\eqref{eq:general_time} one obtains after a complicated calculation a very simple result
\begin{align}
W^{(n)}(t) 
= e^{-iH^\times t} W^{(n)} e^{iHt} 
= e^{-\Gamma t} \sum_{k=0}^n |z^-\succ^{(k)}\,^{(k-n)}\prec^-\!z|
= e^{-\Gamma t} W^{(n)}(0),\quad t\geq 0
\label{eq:wn_time}
.
\end{align}
	This result means that the complicated non-reducible (i.e., ``mixed'') microphysical state operator $W^{(n)}$ 
    defined by~\eqref{eq:density_op} and \eqref{eq:density_def}
    has a simple and purely exponential semigroup time evolution, 
    like the 0-th order Gamow state~\eqref{eq:wg_time}, and thus leads to the exponential law for the probabilities.
	The operators $W^{(n)}$ are probably the only operators formed by the dyadic 
    products $|z_\mathcal{N}^-\succ^{(m)}\,^{(\ell)}\prec^-\!\!z_\mathcal{N}|$ 
    with $m,\ell=0,1,\cdots,n$, that have purely exponential time evolution.
	Thus the operators $W^{(n)}$, $n=1,2,\cdots,\mathcal{N}-1$, are distinguished from all other operators in $\mathcal{M}_{z_\mathcal{N}}$.
	The microphysical decaying state operator associated with the $\mathcal{N}$-th order 
    pole of the unitary S-matrix $W_{PT}$ is the sum \eqref{eq:density_op} of these $W^{(n)}$.
	Because of the result~\eqref{eq:wn_time} (independence of the time evolution of $n$) this sum has again the simple, exponential time evolution
\begin{align}
W_{PT}(t)\equiv e^{-iH^\times t} W_{PT} e^{iHt} = e^{-\Gamma_\mathcal{N} t} W_{PT};\quad t\geq 0
\label{eq:wpt_time}
.
\end{align}

	Thus, the theory that describes exponentially decaying states by Gamow vectors~\eqref{eq:gamow_eigenvalue} also admits the possibility of exponentially decaying states that are associated to S-matrix poles of $\mathcal{N}$-th order at the position $z_\mathcal{N}=E_\mathcal{N}-i\Gamma_\mathcal{N}/2$.
	The ``mixed'' state~\eqref{eq:density_op} associated to the $\mathcal{N}-th$ order 
    S-matrix pole also has an exponential time evolution.
	The probability to register an observable $\Lambda(t)$ (representing, e.g., a detector) 
    in the state $W^{(n)}$ or $W_{PT}$ is obtained using~\eqref{eq:wn_time}~or~\eqref{eq:wpt_time} as:
\begin{align}
Tr(\Lambda(t)W_{PT})
= Tr(e^{iHt}\Lambda e^{-iH^\times t} W_{PT})
= Tr(\Lambda e^{-iHt} W_{PT} e^{iHt})
= e^{-\Gamma_\mathcal{N} t} Tr(\Lambda W_{PT}),
\quad 0<t<\infty
\end{align}
%
	These exponentially decaying states cannot be described by a vector like $\ketbra{\phi^G}{\phi^G}$.
	The simplest choice for this kind of state operator is the one associated to the pole term of a second order S-matrix pole at $z_2$:
\begin{align}
W_{PT} 
&= 2\pi\Gamma (2W^{(0)} - i W^{(1)})
\nonumber\\
&= 2\pi\Gamma (2|z_2^-\succ^{(0)}\,^{(0)}\prec^-\!z_2| - i(|z_2^-\succ^{(0)}\,^{(1)}\prec^-\!z_2| + |z_2^-\succ^{(1)}\,^{(0)}\prec^-\!z_2|))
\end{align}
	In the subspaces $\mathcal{M}_{z_\mathcal{N}}\subset\Phi_+^\times$ associated to the $\mathcal{N}$-th order pole, the Hamiltonian $H$ is non-diagonalizable and so is the state operator.

	The Hardy space theory, which was {\it needed} for the theoretical description of first order pole states (by Gamow vectors), also provides the mathematical means for higher order pole states described by non-diagonalizable operators; this is not possible in the Hilbert space or the Schwartz space.
    This does not constitute a proof that these states exist in nature -- higher order S-matrix
    poles may be excluded for some other physical reasons -- but it provides a mathematical possibility,
    and it is an argument against the exclusion~\cite{ref:goldberger64} of exponentially decaying 
    higher order resonance states.

\ack
We should like to thank the organizers of the 5th International Symposium on Quantum
Theory and Symmetries, and in particular Mariano del Olmo, for the kind
hospitality, and to gratefully acknowledge support from the US National Science Foundation Award No. OISE-0421936.



\begin{thebibliography}{99}
\bibitem{gibbons} Gibbons L K 1993 Ph.D. thesis (University of Chicago: RX-1487)  [see Eq. (2.2)]  Note that this is the decay rate for the $K_S^0$ only, with the interference effects from the $K_L^0$ subtracted out.
\end{thebibliography}
\end{document}